\documentclass[prb,preprint]{revtex4-1} 

\usepackage{amsmath}  
\usepackage{amsfonts} 
\usepackage{graphicx} 

\usepackage[colorlinks=true, pdfstartview=FitV, linkcolor=blue,
citecolor=blue, urlcolor=blue]{hyperref}

\newcommand{\be}{\begin{equation}}
\newcommand{\ee}{\end{equation}}
\newcommand{\bea}{\begin{eqnarray}}
\newcommand{\eea}{\end{eqnarray}}

\begin{document}


\title{Small Oscillations via Conservation of Energy in a Simple Static Equilibrium Problem}

\author{Tia Troy}
\affiliation{Department of Physics, Winona State University, Winona, MN 55987}

\author{Megan Reiner}
\affiliation{Department of Physics, Winona State University, Winona, MN 55987}

\author{Andrew Haugen}
\affiliation{Department of Physics, Winona State University, Winona, MN 55987}

\author{Nathan T. Moore}
\email{nmoore@winona.edu} 
\affiliation{Department of Physics, Winona State University, Winona, MN 55987}


\date{\today}

\begin{abstract}
The work describes an analogy-based small oscillations analysis of a standard static equilibrium lab problem.  In addition to force analysis, a potential energy function for the system is developed, and by drawing out mathematical similarities to the simple harmonic oscillator, we are able to describe (and verify) the period of small oscillations about the static equilibrium state.  The problem was developed and implemented in a standard University Physics course at Winona State University.
\end{abstract}

\maketitle 

\section{Introduction} 
The theory of small oscillations is an oft-overlooked corner of the introductory physics sequence.  Common oscillation examples in the standard texts include the simple harmonic oscillator - a horizontal mass-spring system, and the simple pendulum.  The behavior of both systems can be described with an elementary sinusoid,
\be x(t)=A+B\sin\left(\omega t + \delta \right),
\ee
where $\omega$, the frequency of the oscillation, is a function of some basic system parameters: $\omega=\sqrt{k/m}$ for the mass-spring system, and $\omega=\sqrt{g/l}$ for a pendulum with small enough oscillation amplitude.  

It seems rare  to find an example more complicated that this in the homework sections of introductory books, and this may be due to the difficulty of expressing equations of motion which are both adequate descriptions of the motion of other systems, while still being analytically tractable for first-year students.

Small oscillations are a fundamental feature of nature though, and it seems like a rewarding thing to include this class of motion in the intro course.  In the present work we describe and apply a method for predicting frequency for most any system oscillating in $1D$.  The general approach is to look for mathematical analogies in the functional ``shape" of the system's potential and kinetic energies and then express $\omega$ as a function of specific characteristics of the the system's energies.  We do not believe this is a novel approach to the problem, \cite{bayman_hammermesh}, but the technique does not seem well-known, so we describe it here for a relatively  complicated system of masses and pulleys that could serve as a support for a suspension bridge.  

\section{Example Problem, Static Equilibrium}
\begin{figure}[h!]
\centering
\includegraphics[width=\columnwidth]{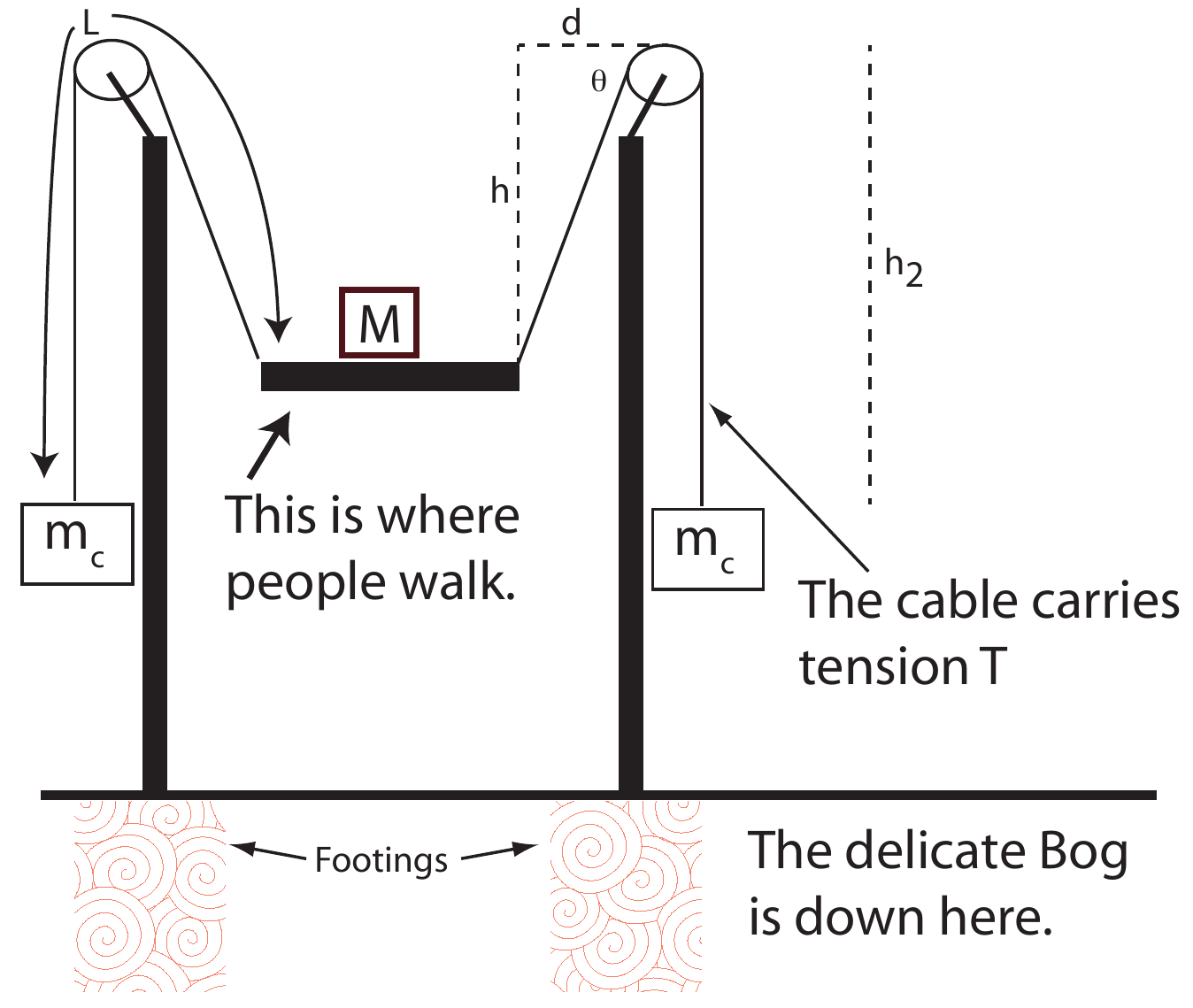}  
\caption{An illustration of the lab setup.  The bridge is made out of a meter stick and several ``knife edge clamps," shown in figure 2. The supporting framework is made out of ring stands and low friction pulleys.  When built, two of these setups could support a section of bridge deck.  For clarity, one might imagine people walking on the bridge in a direction orthogonal to the page. }
\label{fig:overview}
\end{figure}

One of the standard labs in the University Physics sequence at Winona State involves a proposed suspension bridge, illustrated in figure \ref{fig:overview}, which will be used to hold a bridge deck and people above a delicate ecosystem (for example, a bog).  The problem originally comes from the University of Minnesota's Physics department, \cite{UofMN_Lab_manual}, and provides a wonderful setting for introductory students to apply Newton's Second Law to a system in  static equilibrium.  In administering the lab, we normally ask students to predict the bridge's height, $h$, below the top of the pulleys, which normally means they develop a description of the system's equilibrium as a function of applied loads $M$ and $m_c$.  

\begin{figure}[h!]
\centering
\includegraphics[width=\columnwidth]{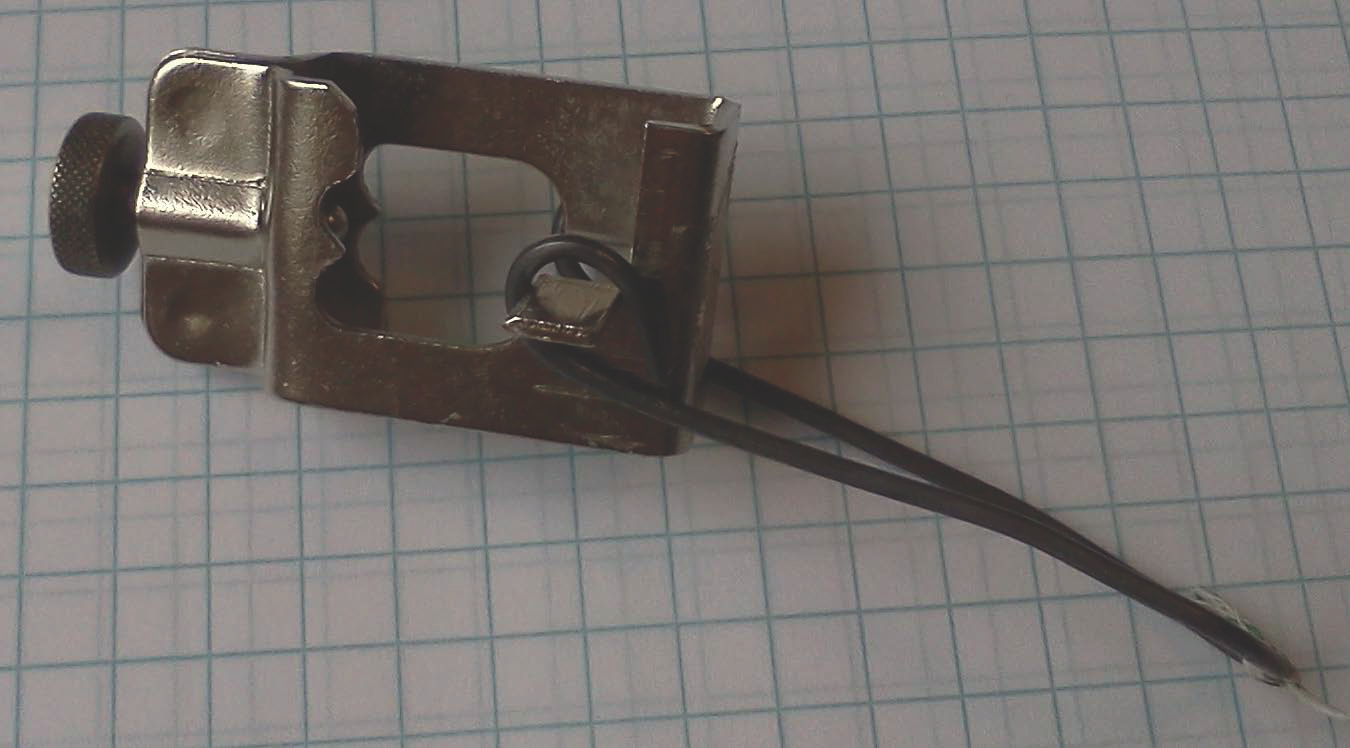}
\caption{
``Knife edge clamps," \cite{clamps}, used for the lab setup.  
}
\label{fig:picture}
\end{figure}
 
A derivation of static equilibrium follows. The central bridge deck and any people on the bridge are described by the net mass $M$.  The counterweights each have mass $m_c$.  The cables and pulleys connecting counterweights to bridge deck are assumed to be frictionless and non-elastic.  The cables carry a tension $T$, and the system has a left-right symmetry (i.e., both cables are at the same angle from the horizontal and carry the same tension).

Balancing vertical forces on either of the counterweights gives 
$T=m_c g$, and balancing forces on the bridge in the vertical direction gives 
$M g = 2T \sin \theta$.
 Pedagogically, it is sometimes useful to ask the students to use the horizontal force balance to prove that if the left-right angles are the same, the left-right cable tensions must also be the same.
 
 These results can be combined to show that in equilibrium,
 \be
 \sin\theta =  \frac{M}{2 m_c}.
 \ee
 Then, looking at the geometry of the system, one can show that if the height of the bridge deck (below the top of the pulleys) is given by $\tan \theta=h/d$, the static equilibrium state of the system, height $h_{eq}$ can be described by
 \be
 \frac{h_{eq}}{d}=\frac{\frac{M}{2m_c}}{\sqrt{1-\left(\frac{M}{2m_c}\right)^2}}.
 \label{eq:equilib}
 \ee
 
\begin{figure}[h!]
\centering
\includegraphics[width=\columnwidth]{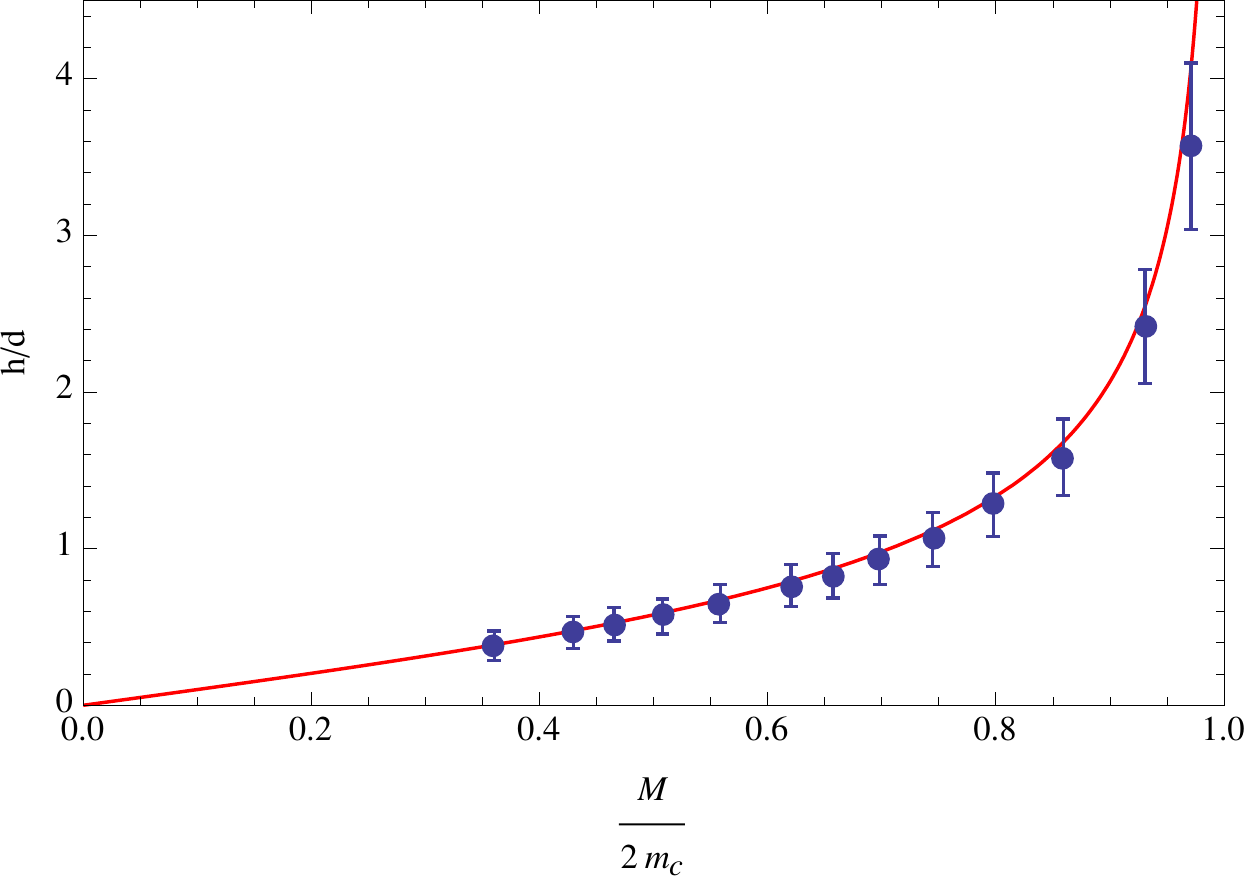}  
\caption{Data taken for the bridge in static equilibrium with $M \approx 220grams$ and $d= 27cm$, uncertainty bars are $2\sigma$. Note the presence of a vertical asymptote for the case of $\frac{M}{2 m_c}\to1$, and further that the non-dimensionalized plot axes allow many different experimental configurations to be shown on the same figure.  Because the bridge deck, $M$, has a minimum value of $\approx220grams$, $\frac{M}{2m_c}$ values of less than $\approx0.35$ are difficult to accomplish.  Similarly, the system lives in a lab with a maximum $h\approx1.2m$, which puts a practical ceiling on the range of $h/d$ values which are accessible.}
\label{fig:static_eq_data}
\end{figure}

Typical student data is compared to this prediction in figure \ref{fig:static_eq_data}.
 
\begin{figure}[h!]
\centering
\includegraphics[width=\columnwidth]{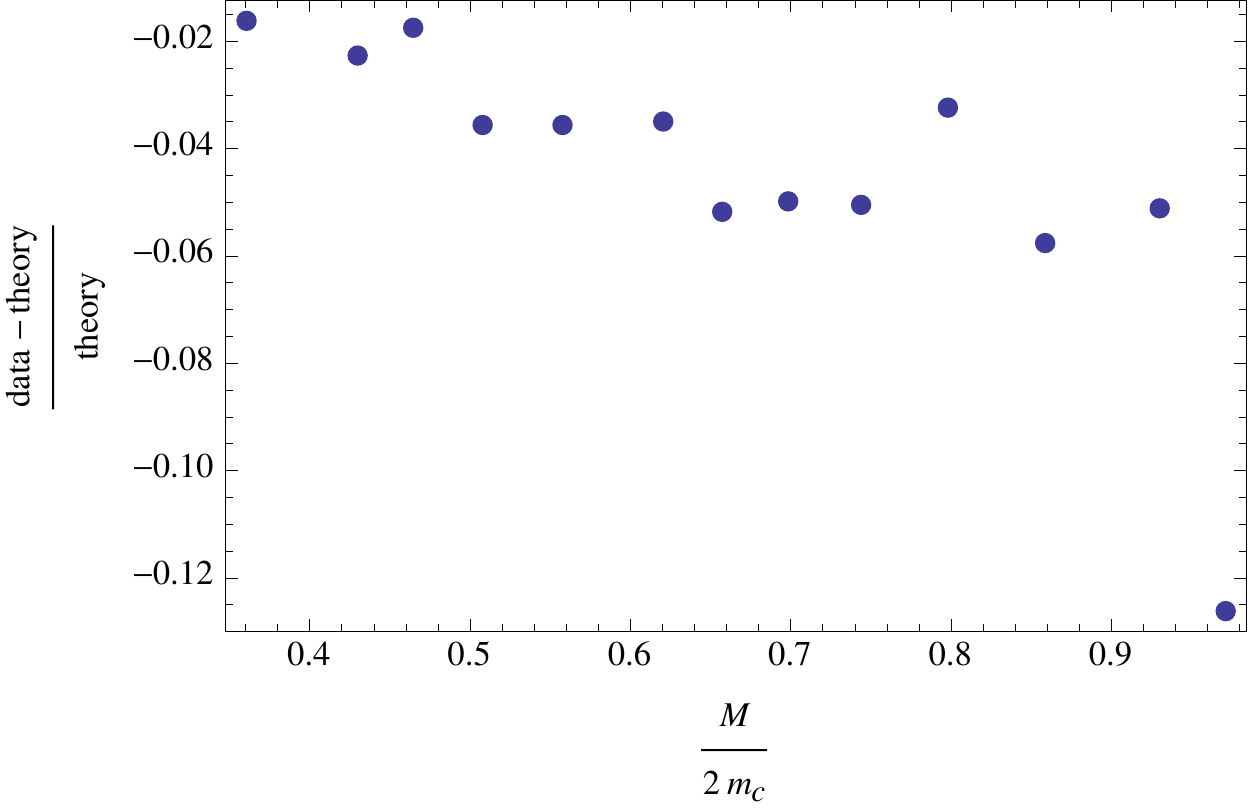}  
\caption{
Residuals showing detail of the $\approx 5\%$ disagreement between data and theory in equation \ref{eq:equilib} and figure \ref{fig:static_eq_data}.  We can ``fix" this by changing to $d \approx 24cm$ in calculations, but that distance doesn't seem to be physically real in the system.  It may be that static friction in the pulleys or knife-edge clamps leads to a dampening term that causes the theory to overpredict $h/d$. 
}
\label{fig:static_eq_residuals_data}
\end{figure} 
 
An equivalent description of equilibrium can be generated with conservation of energy methods.  Specifically, if we ignore the mass of the rotating pulleys, the stored energy will only be gravitational.  The potential energy of the bridge deck can be expressed as $U_M=-M g h$, and that of the counterweights,
\be
U_c=-2 m_c g \left(L-\sqrt{h^2+d^2}\right),
\nonumber
\ee 
where $L$ is the overall length of the cable and the counterweight hangs a distance  $h_2=L-\sqrt{h^2+d^2}$ below the top of the pulley (see figure \ref{fig:overview}).  Note, these potentials are written in ``credit-card" style, in which a mass has no potential energy if it is at the top of the pulley, and normally has a negative potential energy value.

Thus the total potential energy of the system is 
\be
U=-M g h - 2 m_c g \left(L-\sqrt{h^2+d^2}\right).
\ee
If one describes static equilibrium as the minimum of potential energy, $\frac{\partial U}{\partial h}=0$, equation \ref{eq:equilib} can be recovered.  By similar methods, and again ignoring the pulleys and cable, one can show that the kinetic energy of the system is,
\be
K=\frac{1}{2}\left( M  +  2m_c  \frac{h^2}{h^2+d^2} \right) \dot{h}^2.
\ee
Note, a more complete form of the kinetic energy would include rotational kinetic terms corresponding to the 2 pulleys.  As the pulleys are quite light we have opted to neglect these terms.

\section{Example Problem, Small Oscillations}

The lab described is a challenging experience for first and second year students in University Physics.  Taking accurate system measurements can be tricky, and asking students to develop the prediction, equation \ref{eq:equilib}, can be a rather severe acid test of a student's understanding of Newton's second law.  In taking measurements, students often notice that the system oscillates around the equilibrium position, and the rest of the paper describes a way to predict and measure the period of these small oscillations.

First, from a force perspective, in the vertical ($h$) direction, taking down to be $+\hat h$, the forces on the bridge deck are,
\bea
\sum F_h &=& Mg - 2 \overbrace{m_c g}^\text{tension} \sin\theta, \nonumber \\
&=& Mg - 2 m_c g \frac{h}{\sqrt{h^2+d^2}},\\
&=& Mg - 2 m_c g \frac{1}{\sqrt{1+\left(\frac{d}{h}\right)^2}}.
\eea   
Equilibrium is reached when these two terms balance, and if one imagines small deviations from equilibrium, e.g. $h=h_{eq}+\delta$, positive values of $\delta$ make the upward, $-\hat h$, pull from the counterweights bigger than the weight of the bridge deck, and the force shows clear ``restoring" behavior.  However, writing down $\sum F=ma$ for this system to generate the equation of motion seems  beyond the ability of most university physics students, perhaps because of the non-linear relationship between the different masses' accelerations.

Alternately, one can develop the equation of motion for the bridge-counterweight system by taking the time derivative of the system's energy, under the constraint that energy is not lost from the system, e.g. 
\bea
\frac{d}{dt} E_{sys} &=& \frac{d}{dt}\left(K+U \right) \nonumber \\
0 &=& \frac{d}{dt} \left[ 
\underbrace{
	\frac{1}{2} 
	\overbrace{\left( M  +  2m_c  \frac{h^2}{h^2+d^2} \right)}^\text{effective~mass} 
	\dot{h}^2}_\text{Kinetic~Energy} 
\underbrace{ - M g h - 2 m_c g \left(L-\sqrt{h^2+d^2}\right) }_\text{Potential~Energy} \right]
\label{eq:sys_E}
\eea   
As can be seen, this results in a second-order non-linear differential equation, the solution to which, again, seems beyond the ability of most University Physics students.

To describe oscillations around equilibrium beyond hand-waving, the route we propose, and have used for this lab problem, is to interpret the problem with ``analogies."  In this approach we do not intend to be avant-garde, rather, the idea is to remind students of the energy formulation and oscillation frequency of a simple harmonic oscillator, and then look  for equivalent quantities in the system energy of the bridge, equation \ref{eq:sys_E}.  The general method described here is not novel, and follows that in \cite{bayman_hammermesh}.

The energy of a one dimensional simple harmonic oscillator can be written as,
\be
E_{sys}=\frac{1}{2}m \dot{x}^2 + \frac{1}{2}k(x-x_0)^2,
\ee
where the system oscillates about $x_0$ with an angular frequency $\omega=\frac{2 \pi}{T}=\sqrt{k/m}$.

If one looks at the bridge system's energy as described in equation \ref{eq:sys_E}, it seems possible that the bridge has an effective mass of 
\be
M_{eff}=M  +  2m_c  \frac{h^2}{h^2+d^2}.
\ee
To remove the dependance on the bridge deck's height, it seems most appropriate to evaluate this mass at the equilibrium position, $h_{eq}$, given in equation \ref{eq:equilib}.  This simplifies to,
\be
M_{eff}=M \left( 1+\frac{M}{2m_c}\right).
\ee

The spring constant for the bridge system is a little more involved.  Adhering to the analogy of a harmonic oscillator, if we describe the effective spring constant to be $k_{eff}=\frac{\partial^2}{\partial h^2}U$, again evaluated at the equilibrium position, $h_{eq}$,  we find, 
\bea
\frac{\partial^2 U}{\partial h^2} 
	&=& \frac{2 m_c g}
		{\sqrt{h^2+d^2}}
		\left( 1-\frac{h^2}{h^2+d^2} \right) \nonumber \\
k_{eff} &=& \left. \frac{\partial^2 U}{\partial h^2}\right|_{h\to h_{eq}}
	= \frac{2 m_c g}{d} 
		\left(1-\left(\frac{M}{2m_c}\right)^2 \right)^{3/2}
\eea

\begin{figure}[h!]
\centering
\includegraphics[width=\textwidth]{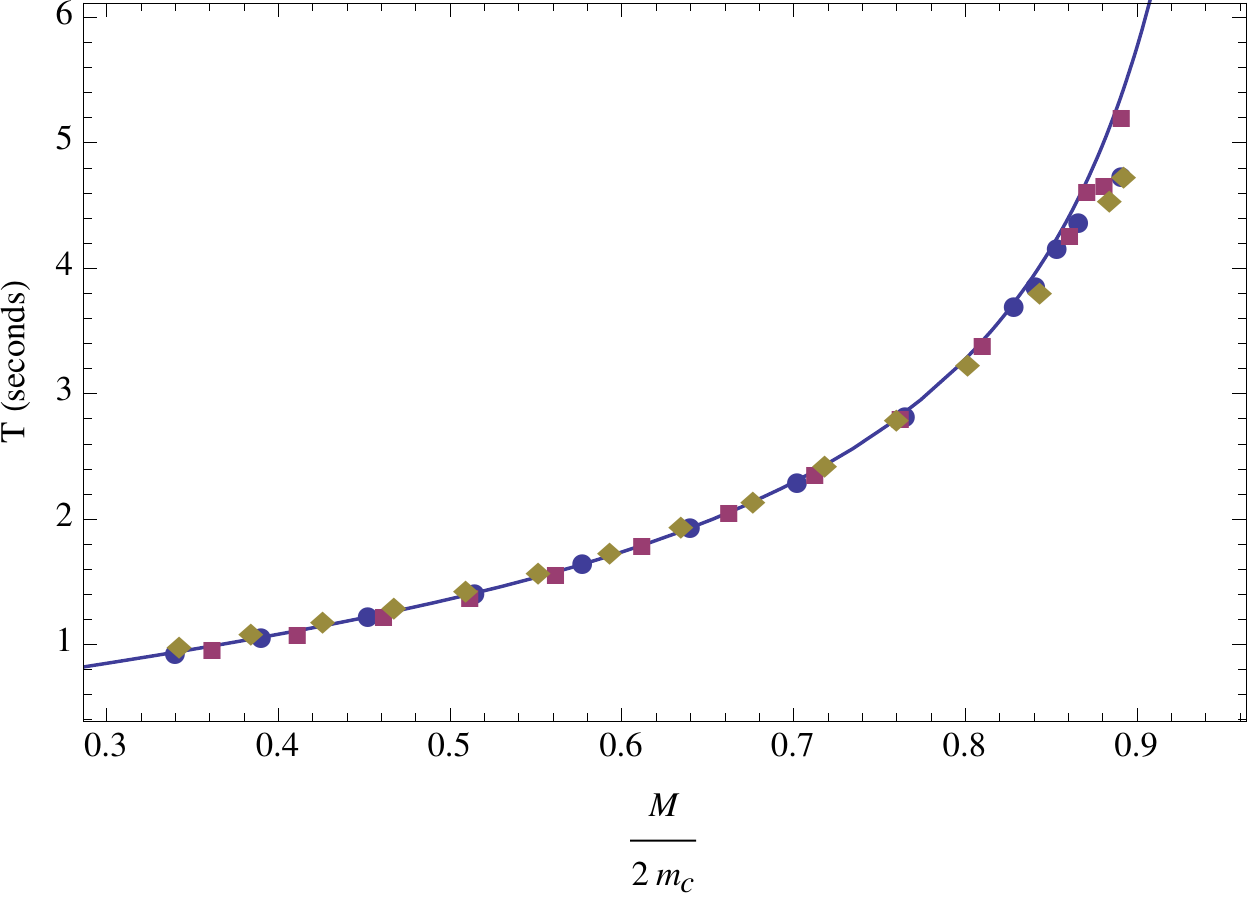}  
\caption{Oscillation period versus $y=\frac{M}{2m_c}$.  
The uncertainty in period measurement is roughly the size of the plot marker.
The data shown was taken at $d=40cm$ and $m_c=[0.2,0.25,0.3]kg$.  For the theory line, equation \ref{eq:period}, we use $g=9.81~N/kg$.  Physically speaking, equilibrium, $h_{eq}$, is poorly defined in the region $\frac{M}{2m_c}>0.8$ and it is correspondingly difficult to get the system to oscillate in a stable, repeatable way in this region.
}
\label{fig:oscillation}
\end{figure}

Given these effective values, we can predict that the bridge will oscillate around equilibrium with angular frequency,
\bea
\omega&=&\sqrt{\frac{k_{eff}}{m_{eff}} } \nonumber\\
&=& \left[ \frac{g}{d} \frac{2 m_c}{M} \frac{(1-(\frac{M}{2m_c})^2)^{3/2}}{1+\frac{M}{2m_c}} \right]^{1/2},~\textrm{or},\nonumber\\
\omega \sqrt{d/g} &=& \left[ \frac{(1-y^2)^{3/2}}{y(1+y)} \right]^{1/2}, \textrm{~where~} y=\frac{M}{2m_c}.
\eea
Similarly, oscillation period $T$ can be predicted,  
\bea
T&=&\frac{2 \pi}{\omega}=2 \pi \sqrt{\frac{m_{eff}}{k_{eff}} } \nonumber \\
T &=& 2 \pi \sqrt{
	\frac{d}{g} 
	\frac	{y(1+y)}
		{\left(1-y^2\right)^{3/2}} 
}, \textrm{~where~} y=\frac{M}{2m_c} \label{eq:period}.
\eea
Data evaluating this prediction is given in figure \ref{fig:oscillation}.

It should be obvious by now that by ``analogy," what we really mean mathematically, is developing a Taylor series expansion for the system energy about equilibrium, $h_{eq}$, and then working with the terms that look like the simple harmonic oscillator to extract an oscillation frequency.

\begin{figure}[h!]
\centering
\includegraphics[width=\textwidth]{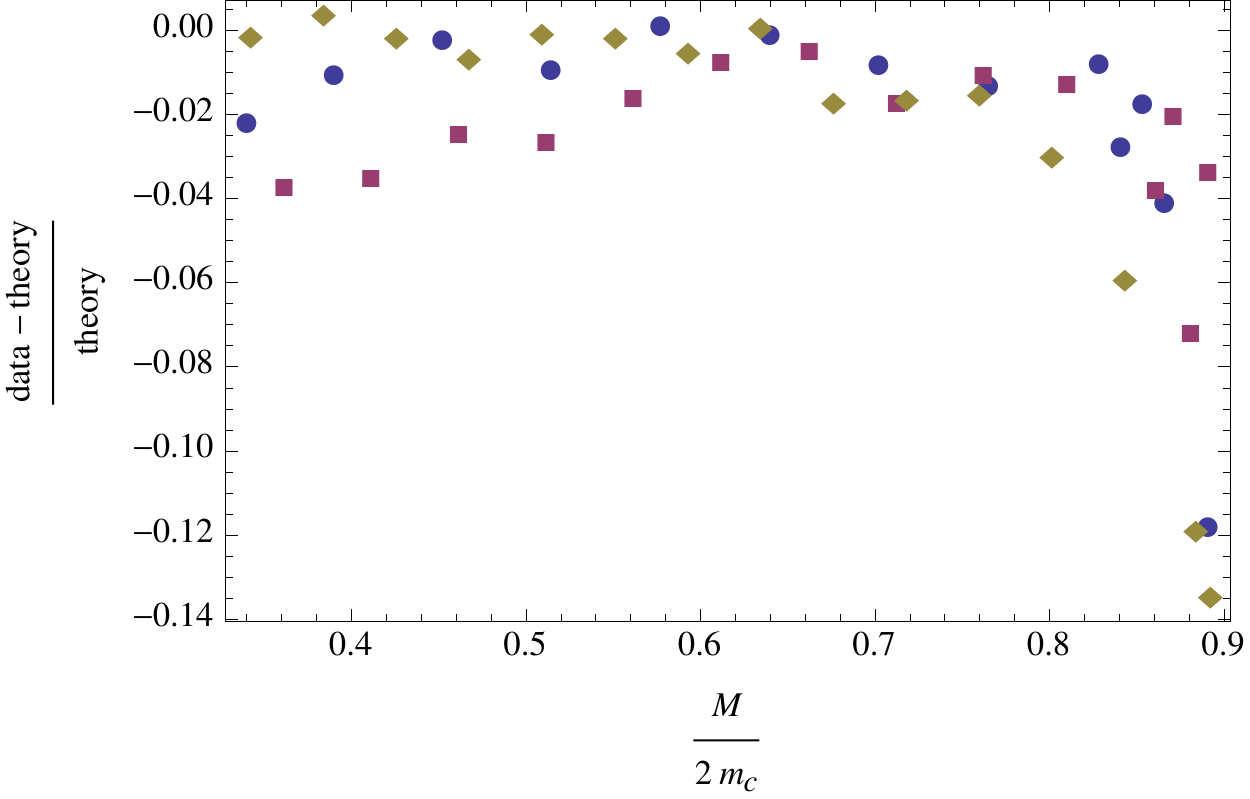}
\caption{
The figure shows residuals in oscillation period between equation \ref{eq:period} and the data in figure \ref{fig:oscillation} above.  The theory seems to describe the system's oscillations to within $5\%$ for $\frac{M}{2m_c}<0.8$.
}
\label{fig:oscillation_residuals}
\end{figure}

\section{Experiment}

Taking repeatable data that agrees with the predictions, equations \ref{eq:equilib} and \ref{eq:period}, is non-trivial.  To measure oscillation period we held $m_c$ at $0.2$, $0.25$, or $0.3$ kg and varied the center mass, $M$, to capture a wide range of $\frac{M}{2m_c}$ values. 
We set the system into motion by displacing the stick from the equilibrium height, $h_{eq}$, and releasing the system with no initial velocity.  To record oscillatory motion we used Vernier's Logger Pro, \cite{Logger_pro_reference}, in conjunction with an ultrasonic distance sensor recording at $50Hz$.

\begin{figure}
\includegraphics[width=\columnwidth]{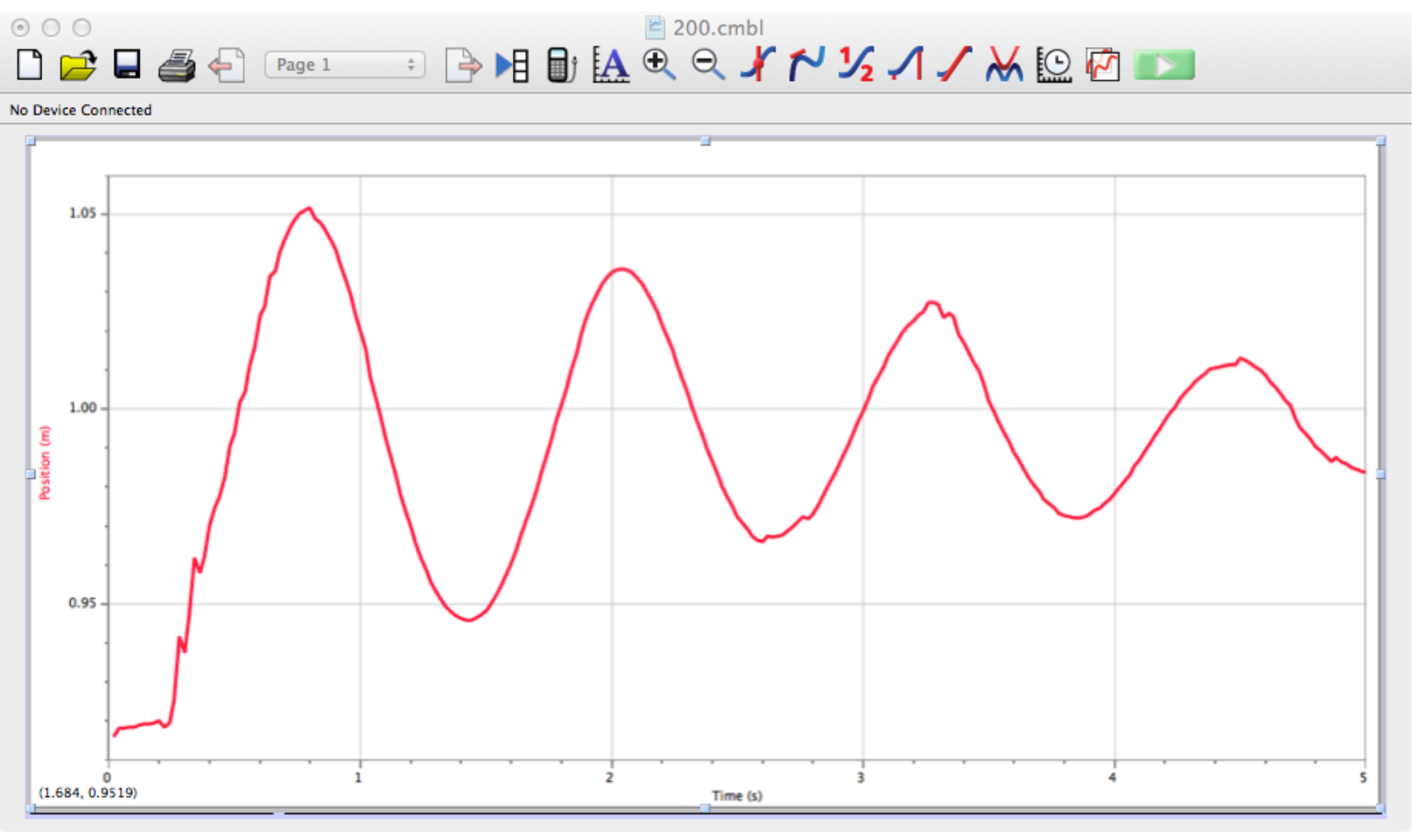}  
\includegraphics[width=\columnwidth]{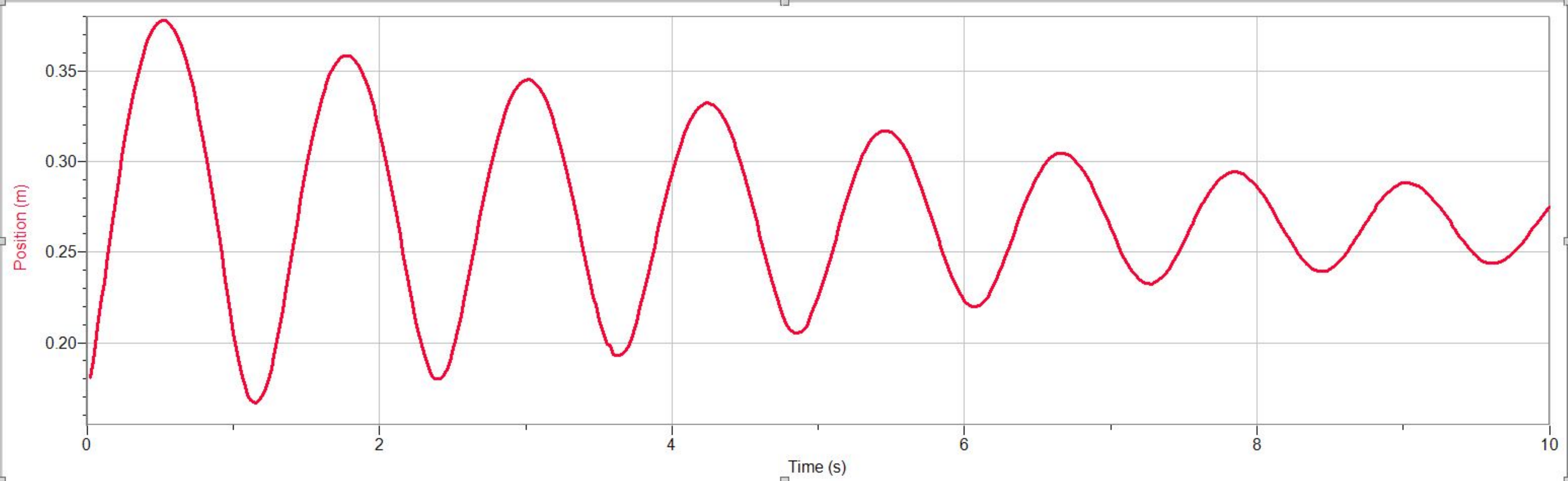}
\caption{Screenshots from LoggerPro of the bridge system's vertical oscillation.  The motion detector is watching from the floor.  The jagged nature of some of the oscillation peaks required fitting each peak with a parabola to determine oscillation period.  Note as well the exponential damping seen in the data.  We are still working to understand the source and nature of damping in the system.}
\label{fig:loggerpro_screenshot}
\end{figure}

\begin{figure}
\includegraphics[width=\columnwidth]{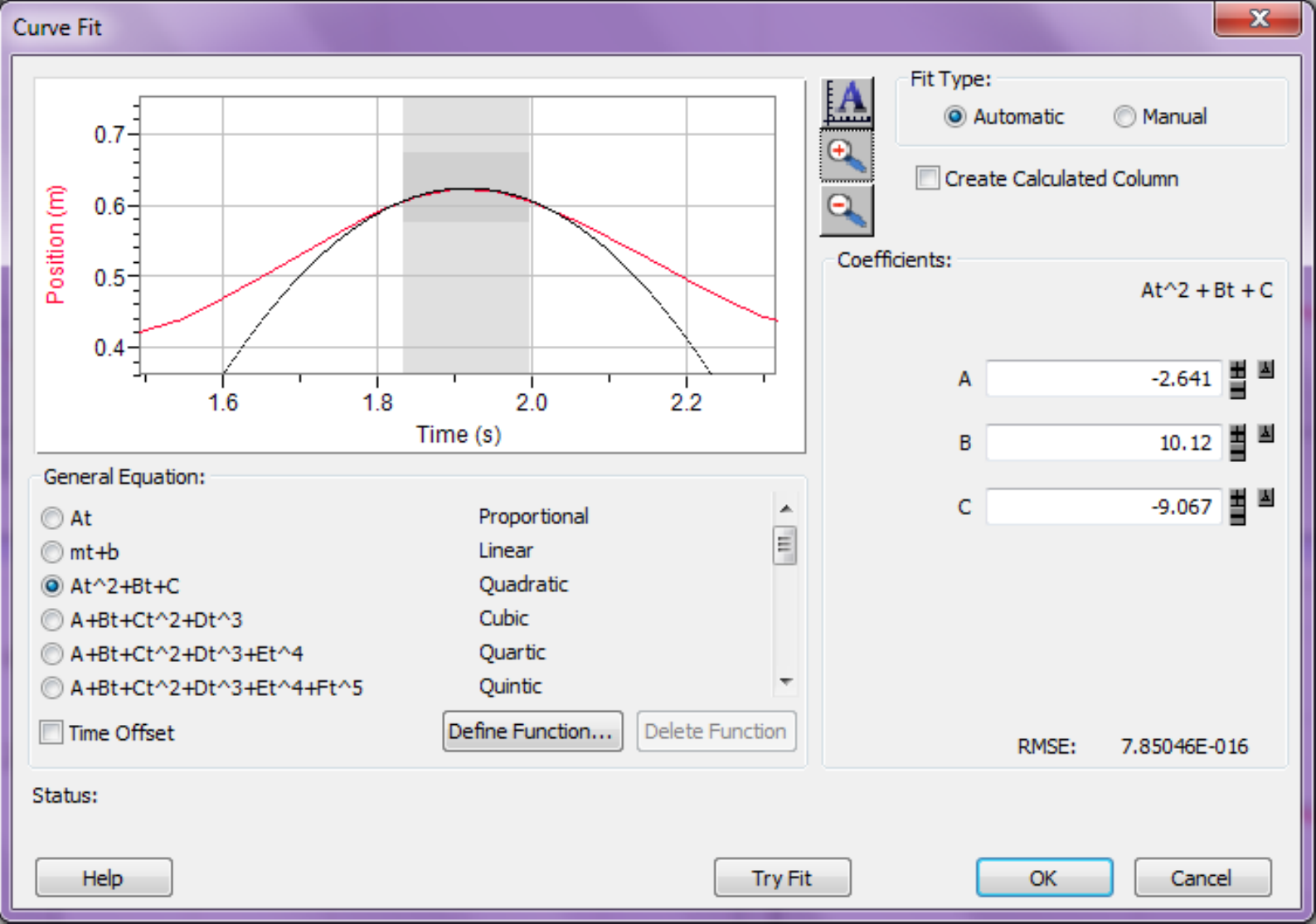}
\caption{Using the method described in equation  in \ref{eq:peak} we determine this oscillation peak to occur at $t=1.916s$. 
}
\label{fig:loggerpro_fit}
\end{figure}

Typical data from LoggerPro is shown in figure \ref{fig:loggerpro_screenshot}.  Some of the oscillation peaks were jagged and we fit parabolas to successive oscillation peaks to more accurately determine the oscillation period.  Specifically, using Logger Pro's quadratic fit we determine the location of the maximum for each crest via (see figure \ref{fig:loggerpro_fit}), 
\bea
f(t)&=&At^2 + Bt + C\nonumber\\
\frac{d f(t)}{dt} &=& 2At + B=0\nonumber\\
t&=& \frac{-B}{2A}. \label{eq:peak}
\eea

Physically speaking, equilibrium, $h_{eq}$, is poorly defined in the region $\frac{M}{2m_c}>0.8$ and it is correspondingly difficult to get the system to oscillate in a stable, repeatable way in this region.  This is seen in the figures. We think that that higher damping within the system, when compared to that at smaller $\frac{M}{2m_c}$ values, is that cause of this instability.  
That said, we do not fully understand the disagreement between data and theory for $\frac{M}{2m_c}>0.8$ and we are tempted to state that this higher damping (neglected in our theory) could cause the period to lengthen in this region.  This damping argument does not seem to be effective though.

Specifically, following reference \cite{Mechanics_reference}, an oscillating system subject to a damping force of $-b\dot{x}$ can be described with the equation of motion,
$\ddot{x} + 2 \beta \dot{x} + \omega_0^2 x = 0$,
where the damping parameter $\beta=b/2m$ and $\omega_0=\sqrt{k/m}$ is the characteristic angular frequency.  
A standard exercise in junior mechanics is to show that this system will oscillate with angular frequency,
\be
\omega_1 = \omega_0 \sqrt{1 - \left( \frac{\beta}{\omega_0} \right)^2 }.
\ee 
In hand-waving terms, as the damping gets stronger, $\beta$ gets larger and $\omega_1$ gets smaller.  As the frequency decreases, period lengthens, which is \textit{exactly opposite} what we see in  the disagreement of data and experiment is  figure \ref{fig:oscillation}.

It may also be the motion in this region is far enough from equilibrium for the simple harmonic oscillator analysis in this paper to be insufficient to describe the motion, in a way similar to how a pendulum released from an angle $90^{\circ}$ from the horizontal will not show a sinusoidal, $\omega=\sqrt{g/l}$ oscillation.

\section{Extensions, Generalization, and Conclusions}
To illustrate the general nature of this approach to small oscillations, consider the simple pendulum with bob mass $m$, pendulum length $l$, and position $\theta$, where $\theta=0$ describes the lowest-energy position.  The pendulum's system energy can be written,
\be
E_{sys}=\frac{1}{2}\left(m l^2 \right) \dot{\theta}^2-mgl\cos \theta.
\ee 
As before, the effective mass can be read off the kinetic energy, 
$m_{eff}=ml^2$, 
and the effective spring constant comes from concavity of the potential energy, evaluated at equilibrium,
\be
k_{eff}=\frac{ d^2 }{ d \theta ^2 } (-m g l \cos \theta) |_{\theta=0}=mgl.
\ee
Together, these results give the familiar form for small oscillations of a pendulum, 
\be
\omega=\sqrt{\frac{k_{eff}}{m_{eff}}} = \sqrt{\frac{mgl}{m l^2}}=\sqrt{g/l}
\ee
We can imagine a similar analysis could be applied to a marble rolling back and forth in a bowl, or a cork bobbing up and down in the water. 

To summarize the approach for determining the period of small oscillations around equilibrium:
\begin{enumerate}
\item The system should be one-dimensional, and should exhibit some oscillation around a (quasi-stable) equilibrium point.
\item The system's energy should be expressible and losses over time should be left out of the expression for 
$E_{sys}=KE+U$.
\item Effective mass can be extracted from kinetic energy by looking for a term of form 
$\frac{1}{2}\left(m_{eff}\right) \dot{x}^2$.
\item ``Spring" or restoring constant comes from concavity of potential energy at the equilibrium point, 
$k=\frac{d^2}{dx^2}U|_{x\to x_{eq}}$.
\item and finally, $\omega=\sqrt{\frac{k_{eff}}{m_{eff}}}$.
\end{enumerate}

We do not provide a proof for this method in the present work, and we would be excited to heard from others what the limits of applicability are for this approach.  Similarly, it would be fascinating to see if damping could be similarly extracted by looking at a system's energy loss over time.   

We thank Jeremy Armstrong for useful discussions, and also the students in the Spring 2013 section of Physics 221 at Winona State University, who worked on this lab with us.

\end{document}